\documentclass[fleqn,usenatbib]{mnras}
\usepackage{newtxtext,newtxmath}
\usepackage[T1]{fontenc}
\usepackage{anyfontsize}
\DeclareRobustCommand{\VAN}[3]{#2}
\let\VANthebibliography\thebibliography
\def\thebibliography{\DeclareRobustCommand{\VAN}[3]{##3}\VANthebibliography}
\usepackage{graphicx}	
\usepackage{amsmath}	
\title[The outbursts of comet 12P/Pons-Brooks]{Mass of particles released by comet 12P/Pons-Brooks during 2023-2024 outbursts}
\author[Maria Gritsevich et al.]{
Maria Gritsevich,$^{1,2,3}$\thanks{E-mail: maria.gritsevich@helsinki.fi}
Marcin Weso\l{}owski,$^{4}$\thanks{E-mail: mwesolowski@ur.edu.pl}
Alberto J. Castro-Tirado$^{5,6}$\thanks{E-mail: ajct@iaa.es}\\
$^{1}$Faculty of Science, University of Helsinki, Gustaf Hallströmin katu 2, FI-00014 Helsinki, Finland\\
$^{2}$Swedish Institute of Space Physics (IRF), Bengt Hultqvists väg 1, 981 92 Kiruna, Sweden\\
$^{3}$Institute of Physics and Technology, Ural Federal University, Mira str. 19, 620002 Ekaterinburg\\
$^{4}$University of Rzesz\'ow, Faculty of Exact and Technical Sciences, Institute of Physics, Pigonia 1 Street, 35-310 Rzesz\'ow, Poland\\
$^{5}$Instituto de Astrofísica de Andalucía (IAA-CSIC), Glorieta de la Astronomía s/n, E-18008, Granada, Spain\\
$^{6}$Ingeniería de Sistemas y Autom\'atica, Universidad de M\'alaga, Unidad Asociada al CSIC por el IAA, Escuela de Ingenier\'ias Industriales,\\ Arquitecto Francisco Pe\~nalosa, 6, Campanillas, 29071 M\'alaga, Spain
}

\date{Accepted XXX. Received YYY; in original form ZZZ}
\pubyear{\the\year{}}
\begin{document}
\label{firstpage}
\pagerange{\pageref{firstpage}--\pageref{lastpage}}
\maketitle
\begin{abstract}
During its most recent return, comet 12P/Pons-Brooks experienced 14 well-documented outbursts, observed between June 13, 2023, and April 2024, at heliocentric distances ranging from $4.26\,$au to $0.85\,$au. After perihelion, two additional outbursts were observed in summer 2024, at heliocentric distances of $1.20\,$au and $2.26\,$au. Using observational data, we developed a numerical model to estimate the mass ejected during these events, focusing on the sublimation of ice through the porous cometary nucleus. The key factors affecting ejected mass estimates are the outburst amplitude and the active surface area during both quiet sublimation and the outburst phases. Pogson's law was used to express outburst magnitude, incorporating scattering cross-sections of cometary agglomerates. The model iteratively determined the mass ejected in observed outbursts, considering various ice types (H$_{2}$O and CO$_{2}$) controlling sublimation activity. Our results indicate that the mass ejected during these outbursts ranged from 10$^{10}$ to 10$^{13}$ kg. Our findings highlight the significant role of surface morphology and thermodynamic conditions in cometary outbursts, providing insights into the mechanisms driving these phenomena and their implications for cometary evolution and dust trail formation. Based on the analysis of observational data, we propose a six-level classification scheme for cometary outbursts.
\end{abstract}
\section{Introduction}
\label{sec:1}
Occasionally, comets undergo sudden and intense surges in brightness and activity, a phenomenon referred to as cometary outbursts. These events involve the rapid release of significant amounts of gas and dust from the nucleus, temporarily amplifying the comet's brightness and apparent size. Cometary outbursts offer a glimpse into the broader context of cosmic evolution. Direct observations reveal these phenomena as spectacular, involving a sudden increase in brightness by at least 1 magnitude. While astronomical data show a wide range of brightness changes in comets, from numerous mini outbursts as seen with comet 67P/Churyumov-Gerasimenko during the \textit{Rosetta} mission \citep{Vi2016} to larger amplitude events like those exhibited by comets 1P/Halley \citep{West1991}, 174P/Echeclus \citep{Skiff2018}, and 29P/Schwassmann-Wachmann \citep{Miles2016a}, the fundamental problem of brightness change, often linked to the thermodynamic evolution of cometary nuclei under rapidly changing conditions, remains unresolved.

\begin{figure*}
\begin{center}
\includegraphics[width=14cm]{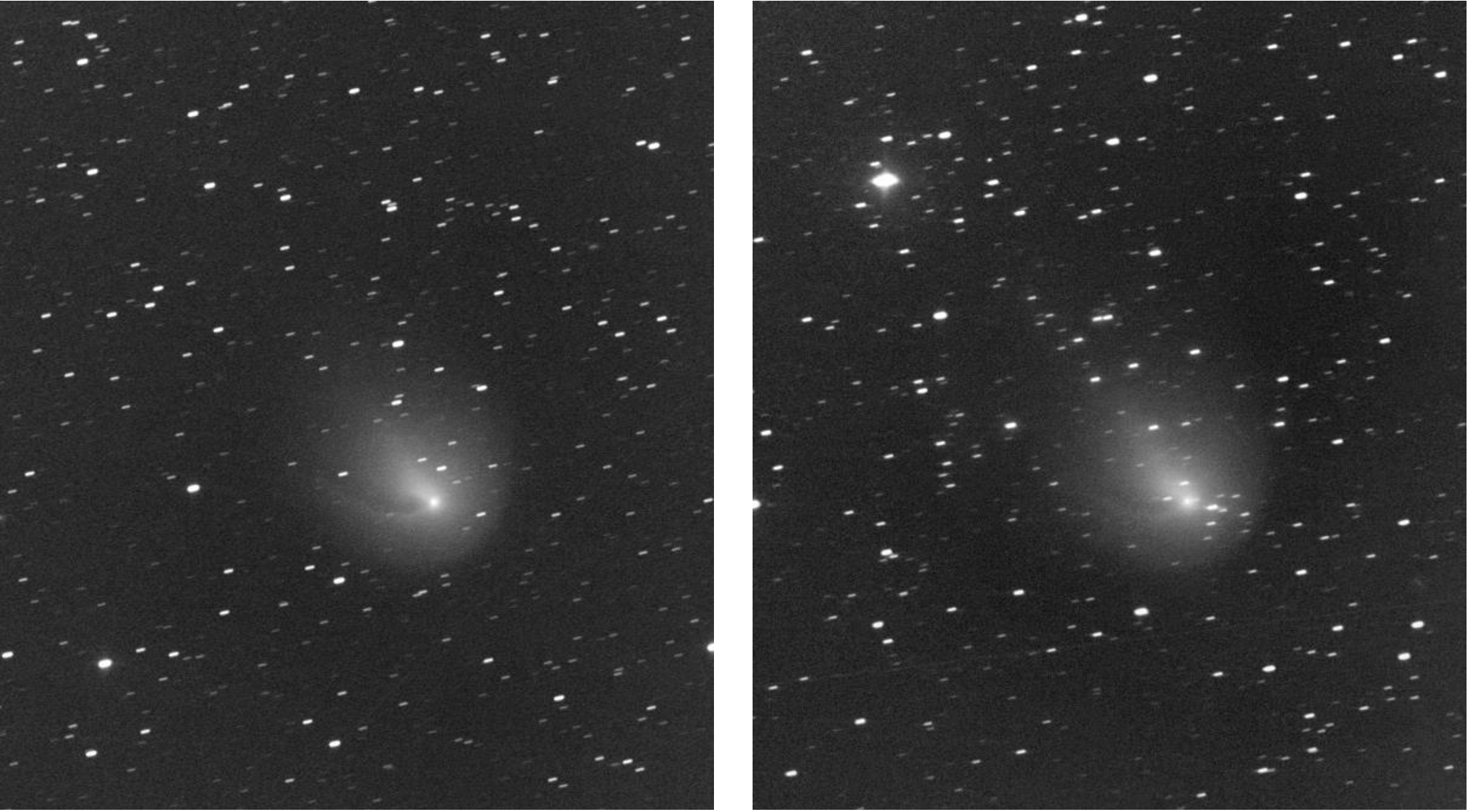}
\end{center}
\caption{Outbursts of comet 12P/Pons-Brooks with a visible coma surrounding its nucleus. Both images were obtained remotely using a T21 0.43-m f/4.5 Corrected Dall-Kirkham astrograph and CCD camera at the iTelescope observatory (U94) in the Great Basin Desert, Beryl Junction, Utah, USA. The total brightness, nuclear brightness, and coma diameter were estimated separately for each date:
1) 2023 November 26.08 UT; m$_{1}$=8.9 mag., m$_{2}$=13.9 mag., Dia.=5.1';
2) 2023 November 27.08 UT; m$_{1}$=8.8 mag., m$_{2}$=13.9 mag., Dia.=5.7'.}
\label{F_0a}
\end{figure*}

Despite extensive research, various mechanisms proposed in the literature \citep{Hughes1990,Prialnik1995,Gronkowski2015,Miles2016a,Miles2016b,Wesolowski2021,Wesolowski2022b,Guliev2022,Ye2022,Belousov2024a,Belousov2024b,Muller2024} have yet to fully elucidate the basic morphological features associated with cometary outbursts. These mechanisms typically involve the destruction of a fragment of the cometary surface, releasing substantial amounts of gas and dust from the nucleus into the coma. This outgassing process contributes to the density increase and expansion of the coma surrounding the nucleus, fundamentally altering the comet's appearance and behavior. The expelled material, composed of volatile gases and dust particles, enhances sunlight scattering on porous dust-ice particles, leading to the characteristic appearance of a bright central nucleus surrounded by a diffuse coma (Fig.\ref{F_0a}), resulting in overall increased comet brightness during an outburst \citep{Wesolowski2021,gritsevich2022}.

The ejection of dust particles and debris from a comet nucleus during outburst events is intricately linked to the formation of meteoroid streams. Appendix A of this paper presents an estimate of the size of particles ejected from the nucleus of comet 12P/Pons-Brooks, necessary for such calculations. Initially appearing to dissipate, these particles form a cloud that gradually expands, primarily influenced by solar radiation pressure and gravitational forces. After half a revolution, the particles reconverge on the opposite side of the Sun near the mutual node of their orbits. Subsequent revolutions bring these particles back to their original outburst location \cite{Lyytinen2013,gritsevich2022}, forming an hourglass-shaped trail due to variations in particle orbits.

The formation of a meteoroid stream is a gradual process spanning multiple orbits of the comet, as its ejected particles follow their orbital trajectories. Over time, these streams can intersect the planetary atmospheres, including those of Earth or Venus, leading to observable meteor showers \citep{Christou2010,Christou2024}. An example of this phenomenon is seen in the connection between comet 12P/Pons-Brooks and the weak December $\kappa$-Draconids meteor shower, typically occurring from November 29 to December 13 \citep{Tomko2016}. 

12P/Pons-Brooks is a periodic comet with an orbital period of approximately 71 years and a nucleus radius of 17$\pm$6 km \citep{Ye2020}. Assuming the density of the cometary nucleus of 500 kg$\cdot$m$^{-3}$, the approximate value of its mass can be estimated 1.03$\cdot$10$^{16}$ kg. This comet is notable for its brightness, reaching an absolute visual magnitude of approximately 5 near perihelion. Initially discovered in July 1812 by Jean-Louis Pons at Marseilles Observatory, it was later observed during its next appearance in 1883 by William Robert Brooks \citep{Yeomans1986}. Ancient records suggest previous apparitions of comet 12P/Pons-Brooks (hereinafter referred to as 12P). With advancements in observational technology over the past 70 years, the comet's most recent perihelion passage on April 21, 2024, and its closest approach to Earth at 1.55 au on June 2, 2024, have been extensively documented. 

In this study, we investigate the initial phases of the well-documented outbursts of comet 12P by employing a numerical model that estimates the ejected mass. Our findings have broader implications for understanding comet behaviour and the formation of dust trails, aiding in predicting the evolution and future observability of such events. Additionally, based on observations, we have developed a classification of outbursts, detailed in section \ref{cldm}. 

\section{Observations of comet 12P/Pons-Brooks}
\label{sec:obs}
\begin{figure}
\begin{center}
\includegraphics[width=8.0cm]{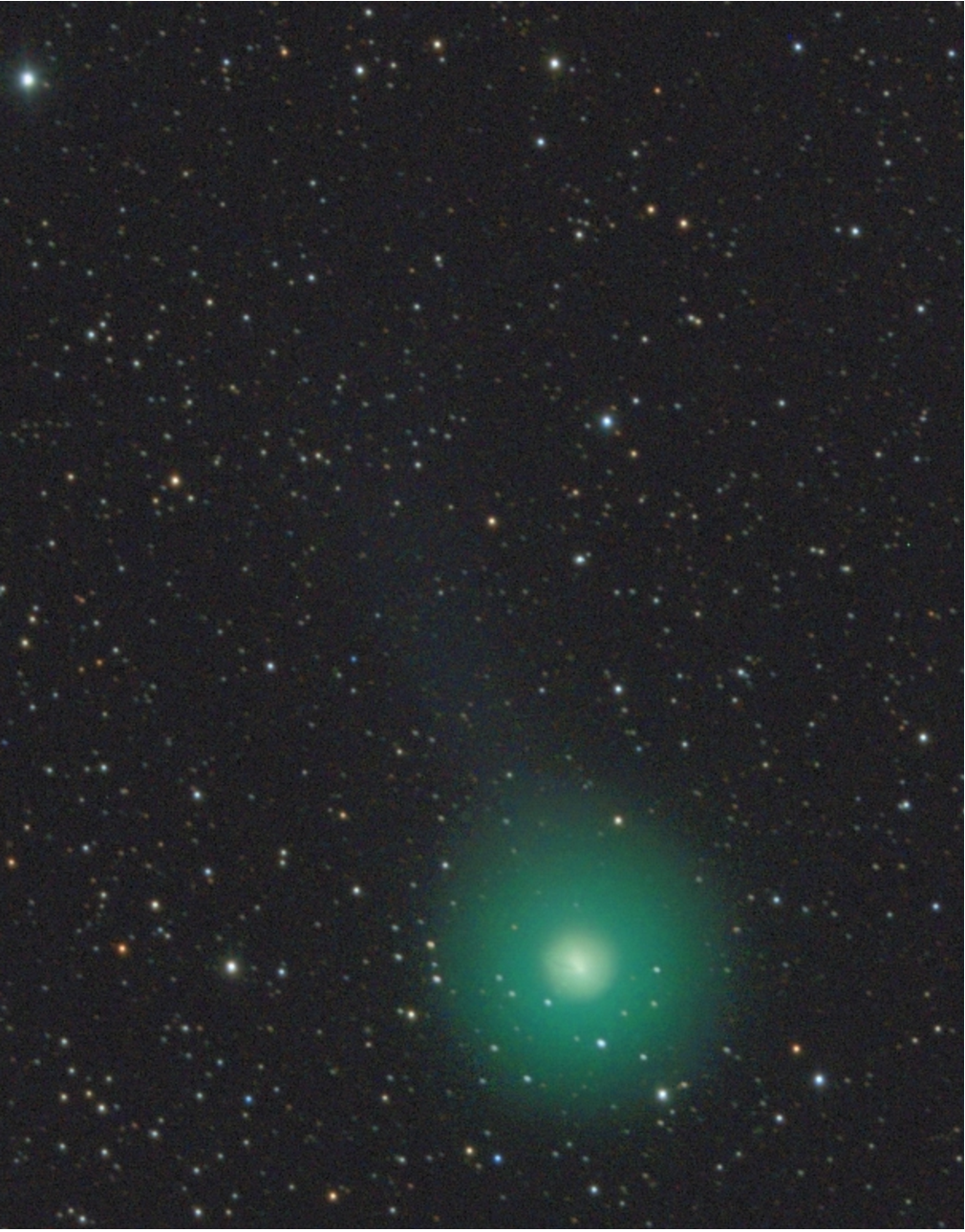}
\end{center}
\caption{Color image of comet 12P/Pons-Brooks, showing the “dark lane” feature visible inside the coma, obtained on 2023 November $19.11\,$UT. The image was captured remotely using a T68 0.28-m f/2.2 RASA astrograph and CMOS camera at the iTelescope observatory (U94) in the Great Basin Desert, Beryl Junction, Utah, USA. The estimated total magnitude is 8.7, with a coma diameter of 9.1 arcmin. A faint tail, approximately 17.7 arcmin in length, is visible at a position angle of 44$^{\circ}$.}
\label{B_1}
\end{figure}

Astronomical observations of comet 12P during its latest return, spanning from the initial signs of activity on June 13, 2023, to April 2024, covered heliocentric distances ranging from $4.26\,$au to $0.85\,$au, and revealed a total of 14 outbursts. After perihelion, two more outbursts were documented at heliocentric distances of $1.20\,$au and $2.26\,$au. The facilities used in early observations included the BOOTES (Burst Observer and Optical Transient Exploring System) — a Global Network of Robotic Astronomical Observatories \citep{1999A&AS..138..583C, 2023NatAs...7.1136C}, the 0.3 m Viestikallio remote observatory in Finland, the remote 0.3 m Makroskooppi observatory in Spain, and remote telescopes at the iTelescope observatories in Utah and California \citep{Prystavski2024,Trigo2024,Borderes2024,Gritsevich2025a}. The expansion rates triggered by the approximately 5-magnitude outburst on 2023-10-05.16 \citep{Usher2023} were earlier reported based on these observations and the initiation time of the outburst \citep{2023ATel16343....1R}. The complete list of instruments used for the observations in this study is presented in Tab.(\ref{Table_0}). Sample images of comet 12P are shown in Figs. \ref{F_0a} and \ref{B_1}; see also  \citep{Gritsevich2025b}. 

Observations of comet 12P continued into 2024, with several outbursts reported before the comet reached perihelion on April 21 and achieved its peak brightness. Despite the increasing brightness, observational conditions became increasingly challenging as the comet approached closer to the Sun. A final pre-perihelion observation was attempted on April 2. At that time, the comet was situated very low in the local evening twilight sky, with an altitude of +16$^{\circ}$ and the Sun at an altitude of -12$^{\circ}$. The comet appeared bright, featuring a long ion tail extending toward the northeast, with a position angle of 39$^{\circ}$. After perihelion, comet 12P/Pons-Brooks reached its closest approach to Earth on June 2, at a distance of $1.55\,$au, becoming more visible from the Southern Hemisphere \citep{Gritsevich2025b}.
\begin{table*}
\caption{List of telescopes and cameras used in observations of comet 12P/Pons-Brooks. }
\label{Table_0}
\begin{tabular}{lllllllll}
\hline
No & Optical Design & Aperture [mm] & F/Ratio & CCD/CMOS& Sensor& Telescope & MPC code\\
 \hline
1. & Hyperbolic Flat-Field Astrograph &	250	& f/3.4	& SBIG ST-10XME	& KAF3200E & T05 & U94 \\
2. & Corrected Dall-Kirkham Astrograph & 610 & f/6.5 & FLI-PL09000 & KAF-09000 & T24 & U69 \\
3. & Corrected Dall-Kirkham Astrograph & 510 & f/4.5 & FLI-PL11002M & KAI-11002 & T11 & U94 \\
4. & Rowe Ackerman Schmidt Astrograph & 279 & f/2.2 & ZWO ASI2600 Color & SONY IMX571 &	T68 & U94\\
5. & Corrected Dall-Kirkham Astrograph & 431 & f/6.8 & FLI-PL16803 & KAF-16803 & T19 & U94 \\
6. & Corrected Dall-Kirkham Astrograph & 431 & f/4.5 & FLI-PL6303E & KAF-6303E & T21 & U94 \\
7. & Petzval Apochromat Astrograph & 106 & f/5.0 & ZWO ASI2400C & Sony IMX410 & T20 & U94 \\
8. & Newtonian & 250 &	f/3.8 &	ASI6200 Pro Series & Sony IMX455 & T75 & X07 \\
\hline 
\end{tabular}
\end{table*}
\section{Methods} 
\label{sec:3}
\noindent Based on the extensive observational data we utilize a numerical model to determine the mass ejected during these outbursts. The key parameter in this determination is the estimation of the ice sublimation flux occurring through the porous structure of the nucleus of comet 12P. Estimating the mass ejected as a result of an outburst is a complex problem that depends on many parameters. In this study, we rely on Pogson's law, which can be defined as:
\begin{equation}
\Delta m = -2.512\mathrm{log}\frac{p(\theta)_{\mathrm{N}} A_{\mathrm{N}} S_{\mathrm{N}} + p(\theta)_{2}\left( C(t_{2}) + C_{\mathrm{ej}} \right)}{p(\theta)_{\mathrm{N}} A_{\mathrm{N}} S_{\mathrm{N}} + p(\theta)_{1} C(t_{1})}.
\label{MW1}
\end{equation}
\noindent In Eq.\ref{MW1} p($\theta$)$_{1}$ represents the phase function of the cometary nucleus, p$(\theta)_{2}$ is the phase function of the cometary agglomerates, A$_{\mathrm{N}}$ is the albedo, S$_{\mathrm{N}}$ is the total area of the comet, C(t$_{1}$) is the scattering cross-section of the cometary agglomerates raised into a coma during the quiet sublimation phase (t$_{1}$), C(t$_{2}$) is the scattering cross-section of the cometary agglomerates raised into a coma during the outburst phase (t$_{2}$), C$_{\mathrm{ej}}$ is the scattering cross-section of agglomerates originating from the destroyed and ejected layer of the nucleus. The individual scattering cross-sections that appear in Eq.\ref{MW1} can be expressed as \citep{Wesolowski2022a}:
\begin{equation}
C(t_{1})= \eta_{1} \xi \frac{\int_{r_{\mathrm{min}}}^{r_{\mathrm{max}}} Q_{\mathrm{scat}}(r) r^{2-q}dr}{\int_{r_{\mathrm{min}}}^{r_{\mathrm{max}}}r^{3-q}dr},
\label{MW2}
\end{equation}
\begin{equation}
C(t_{2})= \left(\eta_{1} + \Delta\eta\right) \xi \frac{\int_{r_{\mathrm{min}}}^{r_{\mathrm{max}}} Q_{\mathrm{scat}}(r) r^{2-q}dr}{\int_{r_{\mathrm{min}}}^{r_{\mathrm{max}}}r^{3-q}dr},
\label{MW3}
\end{equation}
\noindent and 
\begin{equation}
C_{\mathrm{ej}} = \frac{3 M_{\mathrm{ej}}\int_{r_{\mathrm{min}}}^{r_{\mathrm{max}}} Q_{\mathrm{scat}}(r) r^{2-q}dr}{4 \rho_{\mathrm{gr}}(1 - \psi)\int_{r_{\mathrm{min}}}^{r_{\mathrm{max}}}r^{3-q}dr}. 
\label{MW4}
\end{equation}
\noindent In Eqs.(\ref{MW2}-\ref{MW4}) $\eta_{1}$ is the active surface during quiet sublimation, $\Delta\eta$ is a correction related to the ejection of a fragment of the cometary nucleus surface during the outburst, $\xi$ is a factor related to thermodynamic parameters, $r$ is the radius of porous particles on which the incident sunlight scatters, Q$_{\mathrm{scat}}$(r) is the scattering coefficient, $q$ is an index in the power law, M$_{\mathrm{ej}}$ is the mass ejection, $\rho_{\mathrm{gr}}$ is the density of particles, and $\psi$ is the porosity of particles. Eqs.(\ref{MW2}-\ref{MW3}) contain two parameters that can be expressed as:
\begin{equation}
\xi = \frac{3 S_{\mathrm{N}}\gamma_{\mathrm{j}}\psi F_{\mathrm{i}} R_{\mathrm{c}}}{v_{\mathrm{g}} \rho_{\mathrm{agg}}},
\label{MW5}
\end{equation}
\noindent and 
\begin{equation}
\Delta\eta = \frac{M_{\mathrm{ej}}}{4 S_{\mathrm{N}} h \rho_{\mathrm{gr}}(1 - \psi)}.
\label{MW6}
\end{equation}
\noindent In Eqs.(\ref{MW5}-\ref{MW6}) $\gamma_{\mathrm{j}}$ is the dust-gas mass ratio (if j = 1 then comet activity occurs in the quiet sublimation phase and if j = 2 then comet activity occurs in the outburst phase), $F_{\mathrm{i}}$ is the sublimation flux (if i=1, this sublimation activity is controlled by ice H$_{2}$O, and when i=2, this sublimation activity is controlled by ice CO$_{2}$), R$_{\mathrm{c}}$ is radius of the cometary coma, $\rho_{\mathrm{agg}}$ is the density of agglomerates, v$_{\mathrm{g}}$ is the gas velocity, and $h$ is the thickness of the destroyed layer. The parameter $\eta(t_{1})$ defines the fraction of the nucleus surface actively sublimating during periods of quiet activity, relative to its total surface area. In our calculations, we consider a broad range for this parameter, extending up to 50\%, reflecting observations from comet 67P/Churyumov-Gerasimenko where water ice sublimation predominantly occurred on sunlit areas of the nucleus   \citep{Gicquel2016}.  
To determine the value of the mass expelled during the outburst, we also need to ascertain individual thermodynamic parameters such as gas velocity and sublimation flux. This requires resolving the energy balance equation, which can be represented as follows: 
\begin{equation}
\frac{S_{\odot}(1 - A_{\mathrm{N}})}{r_{\mathrm{h}}^{2}} \, \mathrm{max}(\mathrm{cos}\,\Theta_{\mathrm{Sun}}, 0)=
\epsilon \,\sigma_{\mathrm{B}} \,T^{4}_{\mathrm{i}} + H_{\mathrm{k}}\,F_{\mathrm{i}}, 
\label{MW7}
\end{equation} 
\noindent where S$_{\odot}$ is the solar constant at $1\,$au, $\Theta_{\mathrm{Sun}}$ is the solar zenithal angle, r$_{\mathrm{h}}$ is the heliocentric distance at which the comet outburst was observed, $\epsilon$ is the emissivity, $\sigma_{\mathrm{B}}$ is the Stefan Boltzmann constant, T is the temperature, and $H_{\mathrm{k}}$ is the latent heat of sublimation (if k=1 then we take into account the latent heat of sublimation of ice H$_{2}$O, and if k=2 then we take into account the latent heat of sublimation of ice CO$_{2}$). Let us explain that correctly describing the sublimation of cometary ice is not a simple task. Using thermodynamic models, one can determine the temperature, but this requires adopting certain values in the model that are poorly defined. An example of such a parameter is the thermal conductivity of the surface layer of the core and the changes occurring in this layer under the influence of the sublimation flux. Therefore, in order not to underestimate or overestimate the temperature value due to the uncertainty of thermal conductivity in the energy balance equation, this factor was omitted.

Additionally, the gas velocity and sublimation flux are calculated using the following relationships:
\begin{equation}
v_{\mathrm{g}} = \sqrt{\frac{\pi k_{\mathrm{B}} T_{\mathrm{i}}}{2 m_{\mathrm{g,i}}}}, 
\label{MW8}
\end{equation}
\noindent and 
\begin{equation}
F_{\mathrm{i}} = \beta \psi p_{\mathrm{sat,i}} \sqrt{\frac{\pi m_{\mathrm{g,i}}}{2 k_{\mathrm{B}} T_{\mathrm{i}} }},
\label{MW9}
\end{equation}
\noindent where k$_{\mathrm{B}}$ is the Boltzmann constant, m$_{\mathrm{g,i}}$ is the mass of a gas molecule, $\beta$ is the sticking coefficient of the gas molecules on to the surface (the value of this coefficient is in the range 0$<\beta<$1), and p$_{\mathrm{sat,i}}$ is the pressure of the phase equilibrium. 
\section{Results} 
\label{sec:4}
To determine the actual ejected mass responsible for the outburst, Eq.\ref{MW1} needs to be solved numerically. Put simply, we are seeking the mass ejected value corresponding to a specific outburst amplitude.   Eq.\ref{MW1} is a complex logarithmic function of the individual scattering cross-sections. Hence, we utilized an algorithmic approach for numerical solution. This involved iteratively refining the mass ejected value until convergence criteria were met. In our context, this iterative process identifies the mass ejected value for the observed outburst occurring at a particular heliocentric distance. 

\begin{figure}
\begin{center}
\includegraphics[width=8.5cm]{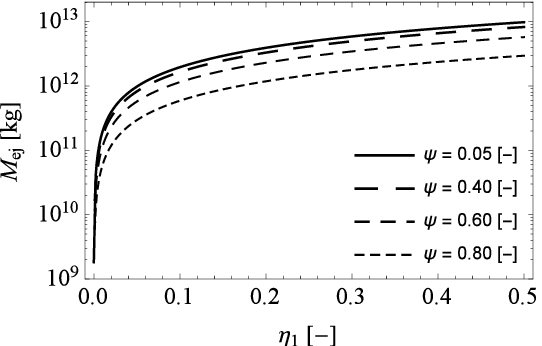}
\end{center}
\caption{The amount of mass ejected during the outburst of comet 12P as a function of the fraction of the active surface during quiet sublimation. In the calculations, it was assumed that the scattering of incident sunlight occurs on porous dust agglomerates with an average radius of r$_{\mathrm{agg}}$ = 1 mm, and the cometary activity was controlled by the sublimation of water ice. These calculations concern the outburst with the largest amplitude $\Delta$m = 5.00 magnitude, which took place at a heliocentric distance of r$_{\mathrm{h}}$ = $2.59\,$au.}
\label{F_1}
\end{figure}

The list of the most important physical constants used in the model is presented in Tab.(\ref{Table_1}). The calculation results of thermodynamic parameters and mass ejected for two example values of the $\eta$ parameter are presented in Tab.(\ref{Table_2}). An example of a graphical interpretation of the mass ejected as a function of the $\eta$ parameter for two extreme values of the outbursts of the comet 12P is presented in Fig.(\ref{F_1}). 

\begin{figure}
\begin{center}
\includegraphics[width=8.5cm]{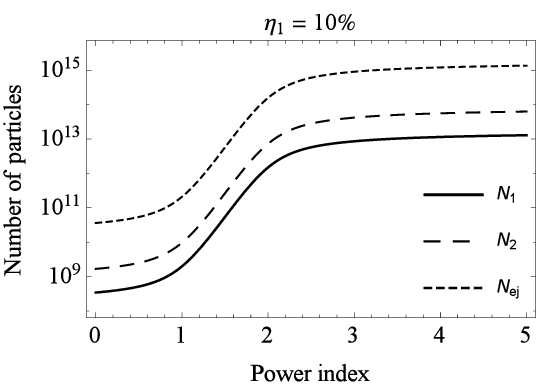}
\includegraphics[width=8.5cm]{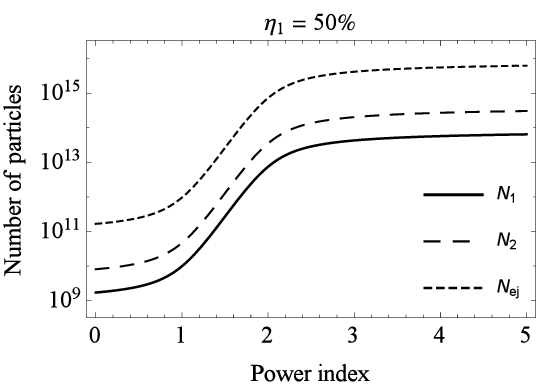}
\end{center}
\caption{The number of particles on which incident sunlight is scattered as a function of the power index q. The considerations took into account that the cometary outburst amplitude was $\Delta$m = 5.00 magnitude, and the outburst took place at a distance of r$_{\mathrm{h}}$ = $2.59\,$au on 2023/07/20.37$\pm$0.08 UT. The number of these particles was calculated per unit area. In addition, the calculations take into account two active surfaces in the quiet sublimation phase, the upper panel concerns the surface $\eta_{1}$ = 10\%, and the lower panel concerns the surface $\eta_{1}$ = 50\%. The calculations assume that the incident sunlight is scattered on dense dust particles with a porosity of $\psi$ = 0.05. The symbols used mean: N$_{1}$ is the number of particles lifted into the coma during the quiet sublimation, N$_{2}$  is the number of particles carried into the coma during the cometary outburst, N$_{\mathrm{ej}}$ is the number of particles created as a result of the destruction of a fragment of the cometary nucleus.}
\label{F_1a}
\end{figure}

\begin{figure}
\begin{center}
\includegraphics[width=8.5cm]{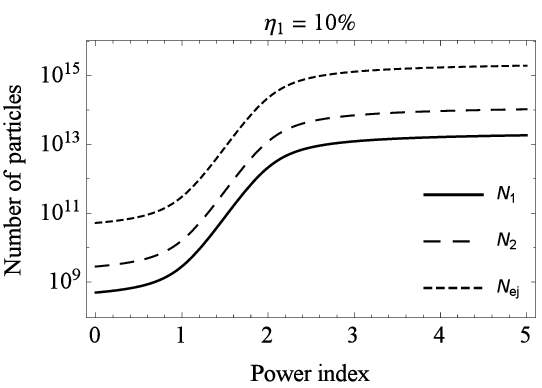}
\includegraphics[width=8.5cm]{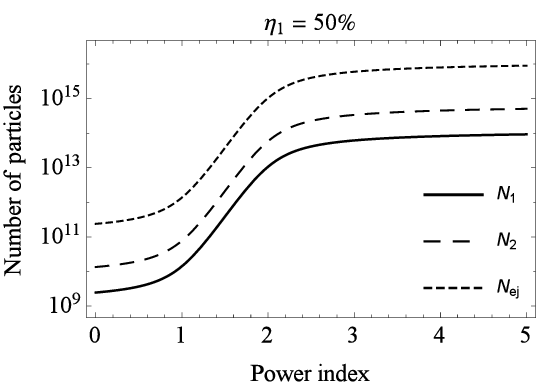}
\end{center}
\caption{The number of particles on which incident sunlight is scattered as a function of the power index q. This situation is analogous to that shown in Fig.(\ref{F_1a}), but the scattering of incident sunlight takes place on dust particles with porosity $\psi$ = 0.8. The remaining symbols are analogous to those shown in Fig.(\ref{F_1a}).}
\label{F_1b}
\end{figure}

\begin{figure}
\begin{center}
\includegraphics[width=8.5cm]{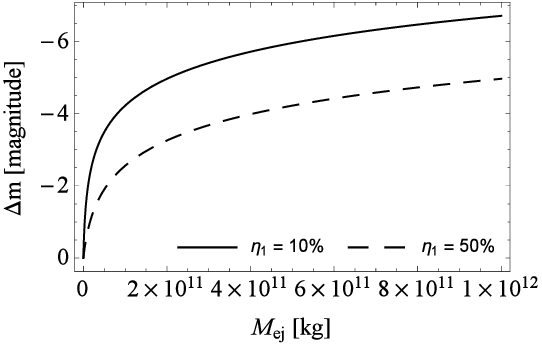}
\includegraphics[width=8.5cm]{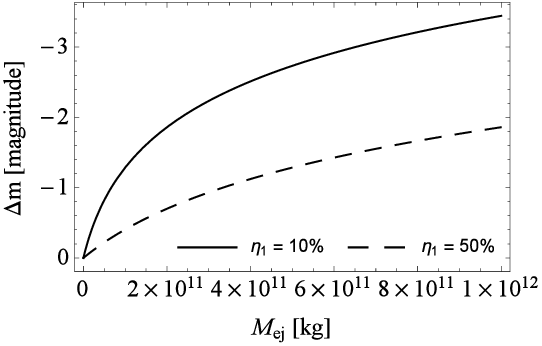}
\includegraphics[width=8.5cm]{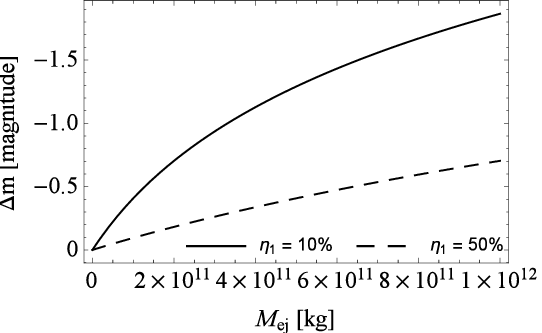}
\end{center}
\caption{The change in comet brightness as a function of the ejected mass for two exemplary values of the active surface area in the quiescent phase of sublimation. The calculations assume that the scattering of incident sunlight occurs on porous dust agglomerates ($\psi$ = 0.7) and that the outburst occurred at a distance of r$_{\mathrm{h}}$ = $3.89\,$au. Furthermore, the upper panel is associated with sublimation controlled by water ice, the middle panel is associated with sublimation controlled by carbon dioxide ice, and the lower panel is associated with sublimation controlled by carbon monoxide ice.}
\label{F6}
\end{figure}

Considering the factors discussed above, we determined the number of particles on which the incident sunlight was scattered, causing the comet to outburst. Key considerations included the ejected mass, which correlates with the active surface area during quiet sublimation, and the power-law distribution used to estimate the average particle size within the coma \citep{Wesolowski2022b}. To illustrate the varying contributions of different cross-sections to the amplitude of the cometary outburst, we quantified the number of particles according to Pogson's law. The results of these calculations are shown in Figs.(\ref{F_1a}-\ref{F_1b}). Moreover, the type of ice responsible for the sublimation activity also influences the amplitude of the cometary outburst. To illustrate this effect, we calculated the outburst amplitude as a function of the ejected mass for two representative values of the active surface during the quiet sublimation phase. The results of these calculations are presented in Fig.\ref{F6}.
\begin{table*}
\caption{The values of cometary parameters used in the numerical simulations.}
\label{Table_1}
\begin{tabular}{lll}
\hline
 Parameter & Value(s)& Reference\\
 \hline
Radius of the cometary nucleus (km) & $R_{\mathrm{N}}$ = 17 $\pm$ 6 & \citep{Ye2020}\\
Albedo of cometary nucleus (-) & $A_{\mathrm{N}}$=0.04& Adopted value\\
Density of cometary agglomerate ($\mathrm{kg \cdot m^{-3}}$) & $\rho_{\mathrm{agg}}$=875 & Adopted value\\
Density of cometary particles ($\mathrm{kg\cdot m^{-3}}$) & $\rho_{\mathrm{gr}}$=2950 &\citep{Davidsson2002}\\
Total area of the comet (m$^{2}$) & S$_{\mathrm{N}}\approx$ 9.08$\cdot$10$^{8}$ & Adopted value\\
Emissivity (-) & $\epsilon$ = 0.9 & Adopted value\\
Radius of the coma (m) & R$_{\mathrm{c}}$=2.5$\cdot$10$^{8}$ & Adopted value\\
Index in the power law (-) & q = 3.5 & Adopted value\\
Average radius of monomers (m) & r = 1.67$\cdot$10$^{-7}$ & Adopted value\\
The upper limit of integration (m) & r$_{\mathrm{max}}$ = 10$^{-2}$ & Adopted value\\
The lower limit of integration (m) & r$_{\mathrm{max}}$ = 10$^{-7}$ & Adopted value\\
The thickness of the destroyed layer (m) & h = 10 & Adopted value\\
Solar constant (for d=1 au) (W$\cdot$m$^{-2}$)& S$_{\odot}$=1361.1& \citep{Gueymard2018}\\
Constant A$_{\mathrm{H_{2}O}}$ for water ice (Pa) & A$_{\mathrm{H_{2}O}}$ = 3.56 $\cdot$ 10$^{12}$&\citep{Prialnik2006}\\ 
Constant B$_{\mathrm{H_{2}O}}$ for water ice (K) & B$_{\mathrm{H_{2}O}}$ = 6141.667&\citep{Prialnik2006}\\ 
Latent heat of water ice sublimation (J$\cdot$kg$^{-1}$) &  H$\mathrm{_{H_{2}O}}$= 2.83 $\cdot$ 10$^{6}$&\citep{Prialnik2006}\\ 
The molar mass of water ice (g$\cdot$mol$^{-1}$) & m$\mathrm{_{H_{2}O}}$ = 18 & Adopted value\\
Constant A$_{\mathrm{CO_{2}}}$ for carbon dioxide (Pa)& A$_{\mathrm{CO_{2}}}$ = 107.9 $\cdot$ 10$^{10}$&\citep{Prialnik2006}\\ 
Constant B$_{\mathrm{CO_{2}}}$ for carbon dioxide (K)& B$_{\mathrm{CO_{2}}}$ = 3148.0 & \citep{Prialnik2006}\\
Latent heat of carbon dioxide sublimation (J$\cdot$kg$^{-1}$)&  H$\mathrm{_{CO_{2}}}$=0.954 $\cdot$ 10$^{6}$& \citep{Prialnik2006}\\ 
The molar mass of water ice (g$\cdot$mol$^{-1}$) & m$\mathrm{_{CO_{2}}}$ = 44 & Adopted value\\
Constant A$_{\mathrm{CO}}$ for carbon dioxide (Pa)& A$_{\mathrm{CO}}$ = 0.1263 $\cdot$ 10$^{10}$&\citep{Prialnik2006}\\ 
Constant B$_{\mathrm{CO}}$ for carbon dioxide (K)& B$_{\mathrm{CO}}$ = 764.16 & \citep{Prialnik2006}\\
Latent heat of sublimation of carbon monoxide (J$\cdot$kg$^{-1}$)&  H$\mathrm{_{CO}}$=0.227 $\cdot$ 10$^{6}$& \citep{Prialnik2006}\\ 
The molar mass of water ice (g$\cdot$mol$^{-1}$) & m$\mathrm{_{CO}}$ = 28 & Adopted value\\
The dust-to-gas mass ratio in the quiet sublimation phase (-) & $\gamma_{\mathrm{1}}$ = 1 & \citep{WesolowskiPotera2024}\\
The dust-to-gas mass ratio in the outburst phase (-) & $\gamma_{\mathrm{2}}$ = 3 & \citep{WesolowskiPotera2024}\\
Wavelength of electromagnetic solar radiation (m)& $\lambda$ = 0.50 $\cdot$ 10$^{-6}$&\citep{Wesolowski2020}\\
The radius of the agglomerate (m)& r$_{\mathrm{agg}}$ = 1$\cdot$10$^{-3}$ & Adopted value\\
Refractive index  for cometary dust particles (-) & n$_{\mathrm{dust}}$ = 1.60 + 0.005$i$&\citep{Wesolowski2020}\\
The scattering coefficient for cometary dust particles (-)& Q$_{\mathrm{dust}}$(r$_{\mathrm{agg}}$) $\approx$ 1&\citep{Wesolowski2020}\\ 
\hline 
\end{tabular}
\end{table*}
\begin{table*}
\caption{Values of thermodynamic parameters and ejected mass describing the outbursts of comet 12P. In the calculation of the ejected mass (M$_{\mathrm{ej}}$), a wide range of active surfaces in the quiet sublimation phase ($\eta$ parameter) was taken into account (M$_{\mathrm{ej}_{1}}$  corresponds to the active surface $\eta$ = 10\%, and M$_{\mathrm{ej}_{2}}$  corresponds to the active surface $\eta$ = 50\%). The values given in the last row of this table apply to situations when the sublimation of CO$_{2}$ ice is responsible for the outburst(*). The following symbols have been adopted in the table: r$_{\mathrm{h}}$ is the heliocentric distance at which the outburst occurred, $\Delta$m is the change in the cometary brightness, $\psi$ is the porosity of the particle on which the incident sunlight is scattering, T is the temperature at the cometary surface, F(T) is the sublimation flux, and v$_{\mathrm{g}}$ is the gas velocity.}
\label{Table_2}
\begin{tabular}{lllllllll}
\hline
Outburst of date&r$_{\mathrm{h}}$&$\Delta$m&$\psi$ & T & F(T)& v$_{\mathrm{g}}$ & M$_{\mathrm{ej}_{1}}$ & M$_{\mathrm{ej}_{2}}$ \\ & [au] &[mag.] &[-] & [K] & [kg$\cdot$m$^{-2}\cdot$s$^{-1}$]& [m$\cdot$s$^{-1}$]&[kg] & [kg]\\
\hline
2023/07/20.37$\pm$0.08 UT & 3.89& 5.50 & 0.40 & 189.294 & 7.347$\cdot$10$^{-6}$& 370.577 & 5.596$\cdot$10$^{11}$&2.798$\cdot$10$^{12}$\\ 
2023/07/20.37$\pm$0.08 UT & 3.89& 5.50 & 0.60 & 187.549 & 8.187$\cdot$10$^{-6}$& 368.865&4.177$\cdot$10$^{11}$ &2.088$\cdot$10$^{12}$\\
2023/07/20.37$\pm$0.08 UT & 3.89& 5.50 & 0.80 & 186.291 & 8.779$\cdot$10$^{-6}$& 367.625&2.247$\cdot$10$^{11}$&1.123$\cdot$10$^{12}$\\ 
2023/09/04.00$\pm$0.60 UT & 3.41& 0.36 & 0.40 & 193.393 & 1.446$\cdot$10$^{-5}$& 374.5672&2.773$\cdot$10$^{9}$&1.386$\cdot$10$^{10}$\\ 
2023/09/04.00$\pm$0.60 UT & 3.41& 0.36 & 0.60 & 191.341 & 1.551$\cdot$10$^{-5}$& 372.575&1.993$\cdot$10$^{9}$&9.964$\cdot$10$^{9}$\\
2023/09/04.00$\pm$0.60 UT & 3.41& 0.36 & 0.80 & 189.887 & 1.623$\cdot$10$^{-5}$& 371.156&1.047$\cdot$10$^{9}$&5.235$\cdot$10$^{9}$\\ 
2023/09/23.87$\pm$0.02 UT & 3.19& 0.90 & 0.40 & 195.158 & 1.918$\cdot$10$^{-5}$& 376.273&1.201$\cdot$10$^{10}$&6.001$\cdot$10$^{10}$\\ 
2023/09/23.87$\pm$0.02 UT & 3.19& 0.90 & 0.60 & 192.992 & 2.032$\cdot$10$^{-5}$& 374.178&8.527$\cdot$10$^{9}$&4.262$\cdot$10$^{10}$\\
2023/09/23.87$\pm$0.02 UT & 3.19& 0.90 & 0.80 & 191.464 & 2.110$\cdot$10$^{-5}$& 372.694&4.445$\cdot$10$^{9}$&2.222$\cdot$10$^{10}$\\ 
2023/10/05.16$\pm$0.03 UT & 3.06& 5.00 & 0.40 & 196.185 & 2.255$\cdot$10$^{-5}$& 377.261&1.063$\cdot$10$^{12}$&5.317$\cdot$10$^{12}$\\ 
2023/10/05.16$\pm$0.03 UT & 3.06& 5.00 & 0.60 & 193.956 & 2.374$\cdot$10$^{-5}$& 375.112&7.508$\cdot$10$^{11}$&3.753$\cdot$10$^{12}$\\
2023/10/05.16$\pm$0.03 UT & 3.06& 5.00 & 0.80 & 192.388 & 2.456$\cdot$10$^{-5}$& 373.593&3.898$\cdot$10$^{11}$&1.949$\cdot$10$^{12}$\\ 
2023/10/22.52$\pm$0.21 UT & 2.86& 0.40 & 0.40 & 197.757 & 2.881$\cdot$10$^{-5}$& 378.770&6.190$\cdot$10$^{9}$&3.094$\cdot$10$^{10}$\\ 
2023/10/22.52$\pm$0.21 UT & 2.86& 0.40 & 0.60 & 195.439 & 3.009$\cdot$10$^{-5}$& 376.544&4.334$\cdot$10$^{9}$&2.167$\cdot$10$^{10}$\\
2023/10/22.52$\pm$0.21 UT & 2.86& 0.40 & 0.80 & 193.814 & 3.095$\cdot$10$^{-5}$& 374.975&2.238$\cdot$10$^{9}$&1.119$\cdot$10$^{10}$\\  
2023/10/31.46$\pm$0.20 UT & 2.76& 2.90 & 0.40 & 198.545 & 2.477$\cdot$10$^{-5}$& 379.524&2.089$\cdot$10$^{11}$&1.044$\cdot$10$^{12}$\\ 
2023/10/31.46$\pm$0.20 UT & 2.76& 2.90 & 0.60 & 196.185 & 3.253$\cdot$10$^{-5}$& 377.262&1.458$\cdot$10$^{11}$&7.288$\cdot$10$^{11}$\\
2023/10/31.46$\pm$0.20 UT & 2.76& 2.90 & 0.80 & 194.533 & 3.472$\cdot$10$^{-5}$& 375.669&7.512$\cdot$10$^{10}$&3.755$\cdot$10$^{11}$\\ 
2023/11/01.40$\pm$0.15 UT & 2.75& 2.50 & 0.40 & 198.624 & 3.293$\cdot$10$^{-5}$& 379.599&1.416$\cdot$10$^{11}$&7.082$\cdot$10$^{11}$\\ 
2023/11/01.40$\pm$0.15 UT & 2.75& 2.50 & 0.60 & 196.261 & 3.424$\cdot$10$^{-5}$& 377.334&9.879$\cdot$10$^{10}$&4.939$\cdot$10$^{11}$\\
2023/11/01.40$\pm$0.15 UT & 2.75& 2.50 & 0.80 & 194.605 & 3.512$\cdot$10$^{-5}$& 375.739&5.089$\cdot$10$^{10}$&2.544$\cdot$10$^{11}$\\ 
2023/11/14.65$\pm$0.05 UT & 2.59& 5.00 & 0.40 & 199.897 & 3.512$\cdot$10$^{-5}$& 380.814&1.867$\cdot$10$^{12}$&9.332$\cdot$10$^{12}$\\ 
2023/11/14.65$\pm$0.05 UT & 2.59& 5.00 & 0.60 & 197.468 & 3.646$\cdot$10$^{-5}$& 378.494&1.295$\cdot$10$^{12}$&6.474$\cdot$10$^{12}$\\
2023/11/14.65$\pm$0.05 UT & 2.59& 5.00 & 0.80 & 195.771 & 3.735$\cdot$10$^{-5}$& 376.863&6.649$\cdot$10$^{11}$&3.324$\cdot$10$^{12}$\\ 
2023/11/30.60$\pm$0.02 UT & 2.39& 3.40 & 0.40 & 201.523 & 5.100$\cdot$10$^{-5}$& 382.359&5.284$\cdot$10$^{11}$&2.642$\cdot$10$^{12}$\\ 
2023/11/30.60$\pm$0.02 UT & 2.39& 3.40 & 0.60 & 199.017 & 5.245$\cdot$10$^{-5}$& 379.975&3.645$\cdot$10$^{11}$&1.823$\cdot$10$^{12}$\\
2023/11/30.60$\pm$0.02 UT & 2.39& 3.40 & 0.80 & 197.268 & 5.343$\cdot$10$^{-5}$& 378.301&1.865$\cdot$10$^{11}$&9.324$\cdot$10$^{11}$\\
2023/12/14.57$\pm$0.11 UT & 2.22& 1.65 & 0.40 & 202.951 & 6.298$\cdot$10$^{-5}$& 383.712&1.066$\cdot$10$^{11}$&5.332$\cdot$10$^{11}$\\ 
2023/12/14.57$\pm$0.11 UT & 2.22& 1.65 & 0.60 & 200.382 & 6.449$\cdot$10$^{-5}$& 381.275&7.328$\cdot$10$^{10}$&3.664$\cdot$10$^{11}$\\ 
2023/12/14.57$\pm$0.11 UT & 2.22& 1.65 & 0.80 & 198.591 & 6.552$\cdot$10$^{-5}$& 379.568&3.739$\cdot$10$^{10}$&1.869$\cdot$10$^{11}$\\ 
2024/01/18.40$\pm$0.05 UT &1.77& 1.90 & 0.40 & 207.082 & 1.140$\cdot$10$^{-4}$& 387.597&2.542$\cdot$10$^{11}$&1.271$\cdot$10$^{12}$\\
2024/01/18.40$\pm$0.05 UT &1.77& 1.90 & 0.60 & 204.345 & 1.157$\cdot$10$^{-4}$& 385.027&1.732$\cdot$10$^{11}$&8.661$\cdot$10$^{11}$\\
2024/01/18.40$\pm$0.05 UT &1.77& 1.90 & 0.80 & 202.443 & 1.168$\cdot$10$^{-4}$& 383.231&8.788$\cdot$10$^{10}$&4.393$\cdot$10$^{11}$\\
2024/02/02.95$\pm$0.75 UT &1.57& 0.60 & 0.40 & 209.169 & 1.525$\cdot$10$^{-4}$& 389.546&5.273$\cdot$10$^{10}$&2.636$\cdot$10$^{11}$\\
2024/02/02.95$\pm$0.75 UT &1.57& 0.60 & 0.60 & 206.355 & 1.543$\cdot$10$^{-4}$& 386.917&3.581$\cdot$10$^{10}$&1.790$\cdot$10$^{11}$\\
2024/02/02.95$\pm$0.75 UT &1.57& 0.60 & 0.80 & 204.401 & 1.555$\cdot$10$^{-4}$& 385.080&1.813$\cdot$10$^{10}$&9.066$\cdot$10$^{10}$\\
2024/02/29.40$\pm$0.20 UT &1.21& 0.70 & 0.40 & 213.595 & 2.773$\cdot$10$^{-4}$& 393.645&1.164$\cdot$10$^{11}$&5.820$\cdot$10$^{11}$\\ 
2024/02/29.40$\pm$0.20 UT &1.21& 0.70 & 0.60 & 210.627 & 2.793$\cdot$10$^{-4}$& 390.901&7.872$\cdot$10$^{10}$&3.936$\cdot$10$^{11}$\\
2024/02/29.40$\pm$0.20 UT &1.21& 0.70 & 0.80 & 208.569 & 2.807$\cdot$10$^{-4}$& 388.986&3.974$\cdot$10$^{10}$&1.987$\cdot$10$^{11}$\\
2024/04/02.95$\pm$0.10 UT &0.85& 0.90 & 0.40 & 219.542 & 5.808$\cdot$10$^{-4}$& 399.088&3.516$\cdot$10$^{11}$&1.758$\cdot$10$^{12}$\\
2024/04/02.95$\pm$0.10 UT &0.85& 0.90 & 0.60 & 216.381 & 5.961$\cdot$10$^{-4}$& 396.204&2.371$\cdot$10$^{11}$&1.185$\cdot$10$^{12}$\\
2024/04/02.95$\pm$0.10 UT &0.85& 0.90 & 0.80 & 214.191 & 5.985$\cdot$10$^{-4}$& 394.194&1.194$\cdot$10$^{11}$&5.972$\cdot$10$^{11}$\\
2024/06/10&1.20& 2.20 & 0.40 & 213.734 & 2.694$\cdot$10$^{-4}$& 393.774&8.581$\cdot$10$^{11}$&4.291$\cdot$10$^{12}$\\
2024/06/10&1.20& 2.20 & 0.60 & 210.762 & 2.825$\cdot$10$^{-4}$& 391.026&5.802$\cdot$10$^{11}$&2.901$\cdot$10$^{12}$\\
2024/06/10&1.20& 2.20 & 0.80 & 208.701 & 2.859$\cdot$10$^{-4}$& 389.109&2.929$\cdot$10$^{11}$&1.464$\cdot$10$^{12}$\\
2024/08/30.39-31.02 UT&2.26& 0.96 & 0.40 & 202.611 & 5.990$\cdot$10$^{-5}$& 383.389&4.048$\cdot$10$^{10}$&2.024$\cdot$10$^{11}$\\
2024/08/30.39-31.02 UT&2.26& 0.96 & 0.60 & 200.056 & 6.140$\cdot$10$^{-5}$& 380.965&2.784$\cdot$10$^{10}$&1.392$\cdot$10$^{11}$\\
2024/08/30.39-31.02 UT&2.26& 0.96 & 0.80 & 198.275 & 6.242$\cdot$10$^{-5}$& 379.265&1.421$\cdot$10$^{10}$&7.107$\cdot$10$^{10}$\\
\hline 
2023/10/05.16$\pm$0.03 UT(*) & 3.06& 5.00 & 0.40 & 108.411 & 1.388$\cdot$10$^{-4}$& 179.373&1.376$\cdot$10$^{13}$&6.884$\cdot$10$^{13}$\\
2023/10/05.16$\pm$0.03 UT(*) & 3.06& 5.00 & 0.60 & 106.903 & 1.392$\cdot$10$^{-4}$& 178.121&9.270$\cdot$10$^{12}$&4.635$\cdot$10$^{13}$\\
2023/10/05.16$\pm$0.03 UT(*) & 3.06& 5.00 & 0.80 & 105.858 & 1.395$\cdot$10$^{-4}$& 177.248&4.667$\cdot$10$^{12}$&2.334$\cdot$10$^{13}$\\
\hline
\end{tabular}
\end{table*}
\section{Classification of cometary outbursts}
\label{cldm}
A cometary outburst is primarily characterized by its brightness amplitude, which represents the magnitude of the observed brightness change. The amplitudes of cometary outbursts can range from subtle variations, as frequently observed in comet 67P/Churyumov–Gerasimenko, to dramatic phenomena like the spectacular October 2007 event of comet 17P/Holmes \citep{Moreno2008,Montalto2008,Trigo2008,Wesolowski2018}.

Based on the amplitude of brightness change during an outburst, we propose a six-level classification scheme (Table~\ref{Tabdm}). The classification categories span from minor glow variations to rare mega-outbursts, using amplitude as the defining metric. This framework enables systematic comparison and analysis of cometary activity across different comets and epochs.

In this study, we applied the classification to sixteen outbursts of comet 12P, finding that the majority of these events were classified as either Class F (six instances) or Class D (six instances), with the remaining two events identified as Class E. This distribution highlights the diverse nature of cometary activity and the prevalence of relatively low-intensity outbursts in the observed dataset.
\begin{table}
\begin{center}
\caption{A six-level classification scheme of cometary outbursts based on observational data.}
\label{Tabdm}
\vspace{0.5cm}
\begin{tabular}{ccc}
\hline
Class & Title & Amplitude [magnitude]\\
 \hline
A & Mega-outburst & 12 $\leq \Delta$m $\leq$ 14 \\
B & Strong outburst & 10 $\leq \Delta$m $<$ 12 \\
C & Intense outburst & 6 $\leq \Delta$m $<$ 10 \\
D & Typical outburst & 2 $\leq \Delta$m $<$ 6 \\
E & Mini-outburst & 1 $\leq \Delta$m $<$ 2 \\
F & Glow variation & 0.01 $\leq \Delta$m $<$ 1 \\
\hline 
\end{tabular}
\end{center}
\end{table}

\section{Discussion}
\label{sec:5}
Despite their sporadic nature, cometary outbursts offer unique opportunities to study the underlying processes shaping cometary activity and the formation of dust trails. In this study, we focused on the well-documented outbursts of comet 12P, aiming to understand the mechanisms driving these phenomena, the mass of particles released, and their implications for cometary evolution.
Through the development of a numerical model based on observational data, we aimed to estimate the mass ejected during these outbursts (Tab.\ref{Table_2}, and Fig.\ref{F_1}). The measure of the total mass ejected is the number of particles coming from individual scattering cross-sections. According to our previous assumption, the largest contribution to the particle number comes from the scattering cross-section associated with the destruction of a fragment of the cometary nucleus (Figs.\ref{F_1a}-\ref{F_1b}). The nucleus challenge in such estimation lies in accurately determining the ice sublimation flux occurring through the porous structure of the cometary nucleus and accounting for the intricate interplay between sublimation activity, surface morphology, and thermodynamic conditions within the comet nucleus. By resolving the energy balance equation and employing relationships governing gas dynamics and sublimation processes, we elucidated the mechanisms contributing to mass ejection during outbursts.

Analyzing the effect of sublimation from individual ices on the amplitude of the outbursts reveals an upward trend in amplitude that is consistent with the increase in the mass ejected, for a fixed ejected mass (Fig.\ref{F6}). This trend corresponds to the number of particles on which incident sunlight is scattered. The outburst amplitude also depends on the active surface area, both during the quiet sublimation phase and the outburst, regardless of the type of ice that caused the outburst.
From the obtained calculation results, it can be concluded that the outburst amplitude is larger when the fraction of the active surface is smaller. This implies that the same amount of mass ejected after the destruction of a fragment of the nucleus surface is relatively larger compared to the mass contained in the coma for smaller values of the parameter $\eta_{1}$. For water-ice controlled sublimation, with an ejected mass of the order of 10$^{12}$ kg, the amplitude was $\Delta$m = -6.72 magnitudes for the parameter $\eta_{1}$ = 10\% and $\Delta$m = -4.97 magnitudes for the parameter $\eta_{1}$ = 50\%. In the case of CO$_{2}$ ice controlled sublimation for the same value of mass ejected for the parameter $\eta_{1}$ = 10\% the amplitude value was $\Delta$m= -3.44 magnitude and for the parameter $\eta_{1}$ = 50\% it was $\Delta$m= -1.86 magnitude. Whereas for sublimation controlled by CO ice for the same value of mass ejected for the parameter $\eta_{1}$ = 10\% the value of the amplitude was equal to $\Delta$m= -1.86 magnitude and for the parameter $\eta_{1}$ = 50\% it was $\Delta$m= -0.76 magnitude.
It is noteworthy that the sublimation flux directly influences the brightness change, with the highest flux observed for CO-ice sublimation and the lowest for water-ice sublimation. This suggests that a smaller sublimation flux leads to a larger amplitude for the same parameter $\eta_{1}$ and the same amount of ejected mass. Our previous research supports this conclusion\textbf{ \citep{Wesolowski2023c}. }

\begin{figure}
\begin{center}
\includegraphics[width=8.5cm]{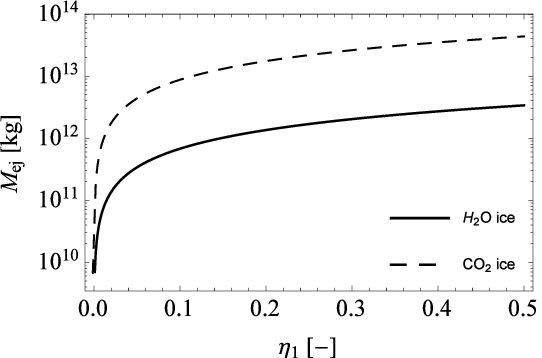}
\end{center}
\caption{Comparison of the amount of mass ejected during the outburst of comet 12P for a distance of r$_{\mathrm{h}}$ = $3.06\,$au as a function of the active surface ($\eta_{1}$). Furthermore, it was assumed that the sublimation activity is controlled by H$_{2}$O ice or CO$_{2}$ ice.}
\label{F_2}
\end{figure}

The comparison of mass ejection between scenarios where sublimation activity is controlled by different types of ice, such as H$_{2}$O and CO$_{2}$, underscores the importance of understanding the composition and behavior of cometary nuclei. These findings not only enhance our understanding of individual cometary events but also have broader implications for our understanding of cometary evolution and the formation of dust trails.
Analyzing the obtained results, we observe that as the porosity of the agglomerate increases, the value of the mass ejected for a given outburst amplitude is lower, which is a consequence of the density of the agglomerates. In the case of H$_{2}$O ice, the upper limit of the mass ejected for the parameter $\eta$=50\% is of the order 10$^{12}$ kg. However, in the case of an outburst controlled by the sublimation of CO$_{2}$ ice, the upper limit of the mass ejected for the parameter $\eta$=50\% is of the order 10$^{13}$ kg.  A comparison of the amount of mass ejected depending on the type of ice responsible for the initiation of the outburst is shown in Fig.(\ref{F_2}). 
The calculated sublimation flux and gas velocity values provide quantitative measures of mass loss and gas production during outbursts, contributing to our understanding of the mechanisms driving cometary activity and the subsequent evolution of the coma and resulting dust trail. Differential effects of radiation pressure on dust particles lead to changes in their orbital parameters, resulting in the formation of enduring dust trails observable in subsequent comet revolutions.
In the context of visual observations of comets, and especially cometary dust trails, the key parameter is the quality of the night sky, i.e., the level of its pollution with artificial light \citep{Wesolowski2019,Wesolowski2023a}. Despite the increasing level of pollution of the night sky with artificial light, comets are an example of celestial bodies that, with a strong outburst, can be visually observed even with the naked eye \citep{Wesolowski2019}, such as the outburst of comet 17P/Holmes in 2007 \citep{Trigo2008,gritsevich2022} and visually comparable outbursts of 12P studied in this paper.

\section{Conclusions}
This study provides valuable insights into the physical properties and dynamic behavior of comet 12P during its outburst events experienced during the last return. The estimated characteristics and albedo of the nucleus suggest a moderately sized object with typical reflective characteristics for a comet. The recorded outburst dates, heliocentric distances, and amplitudes reveal a pattern of periodic activity, influenced by the changes in the comet's internal structure and variations in solar heating. Thermal properties such as emissivity and coma radius indicate significant thermal activity driven by the sublimation of volatile materials. The porosity and temperature values play crucial roles in controlling sublimation rates and the extent of outburst events.
The sublimation flux and gas velocity calculations offer quantitative measures of mass loss and gas production, enhancing our understanding of cometary activity and subsequent coma and dust trail evolution. The comparison between H$_{2}$O and CO$_{2}$ ice-controlled sublimation scenarios show that CO$_{2}$ sublimation results in a higher mass ejection, underscoring the importance of understanding the composition and behavior of cometary nuclei in predicting outburst magnitudes. Increasing agglomerate porosity reduces the mass ejected for a given outburst amplitude due to the lower density of the agglomerates. 
The differential effects of radiation pressure on dust particles released in these outbursts contribute to the creation of enduring dust trails observable in subsequent comet revolutions. Thus, our study not only deepens our understanding of individual cometary outburst events but also provides broader insights crucial for constraining the formation and long-term evolution of meteoroid swarms. 

\section*{Acknowledgements}

We gratefully acknowledge the many collaborators with whom we exchanged observations and engaged in discussions, especially T. Prystavski, J. Ryske, I. P\'erez-Garc\'ia, M. Nissinen, J. M. Trigo-Rodríguez, E. Peña-Asensio, I. Boaca, M. Husárik, O. Ivanova, A. S\'anchez, and J. M. Llenas. We express our gratitude to the Finnish Geospatial Research Institute and the Academy of Finland for supporting the project no. 325806 (PlanetS), which facilitated the relevant observations and development of the methods presented in this paper. The program of development within Priority-2030 is acknowledged for supporting the research at UrFU. This work received support from the Centre for Innovation and Transfer of Natural Sciences and Engineering Knowledge, University of Rzesz{\'o}w, Poland (RPPK.01.03.00-18-001/10-00), the Spanish Ministry of Science, Innovation and Universities projects No PID2023-151905OB-I00  and PID2020-118491GB-I00, Junta de Andalucía grant P20\_010168, and the Centro de Excelencia Severo Ochoa grant CEX2021-001131-S funded by MCIN/AEI/10.13039/501100011033.

\section*{Data Availability}

The data underlying this study are included within the article.


\appendix
\section{Ejection of porous dust agglomerates in the quiet sublimation phase}
\setcounter{figure}{0}
\renewcommand{\thefigure}{A.\arabic{figure}}
An additional mechanism responsible for ejecting porous agglomerates into the coma is quiet sublimation, which peaks near perihelion. To comprehensively describe the activity of comet 12P, we calculated the minimum and maximum radii of porous agglomerates ejected into the coma. To simplify our considerations, we assumed that all the energy that was absorbed by the nucleus of comet 12P was used for the sublimation of water ice. To accurately model the emission of porous agglomerates during quiet sublimation, we considered two scenarios. The first scenario involves agglomerates on the nucleus surface that are not bound to their surroundings, while the second scenario involves agglomerates that are bound to their surroundings. A detailed discussion of these scenarios was recently presented in \citep{Wesolowski2024}. Then the relationships for the minimum and maximum particle radii are provided by Eqs.(\ref{A1}-\ref{A2}): 
\begin{equation}
r_{\mathrm{min}} = \frac{B_{\mathrm{i}}  S_{\mathrm{c}}^{2}}{0.5\,C_{\mathrm{D}} \pi v_{\mathrm{g}} F_{\mathrm{i}}},
\label{A1}
\end{equation}
\noindent and 
\begin{equation}
r_{\mathrm{max}} = \frac{3\,C_{\mathrm{D}}\, v_{\mathrm{g}}\, F_{\mathrm{i}}}{8 \varrho_{\mathrm{gr}}(1-\psi)\left(g_{\mathrm{c}} - 4 \pi^{2} R_{\mathrm{N}} P^{-2} \right)}. 
\label{A2}
\end{equation}
\noindent In Eqs.(\ref{A1}-\ref{A2}), the individual symbols mean: B$_{\mathrm{i}}$ is a parameter related to the cohesion force (B$_{1}$ = 0.036 N$\cdot$m$^{-1}$ \citep{Kossacki2022}, B$_{2}$ = 0.018 N$\cdot$m$^{-1}$ \citep{Thomas2015}), S$_{\mathrm{c}}$ describes the degree of particle cleanliness, C$_{\mathrm{D}}$ is the modified free-molecular drag coefficient for the spherical body (C$_{\mathrm{D}}$ = 2), $v_{\mathrm{g}}$ is the gas velocity which is calculated based on the Eq.(\ref{MW8}), F$_{\mathrm{i}}$ is the sublimation flux which is calculated based on the Eqs.(\ref{MW9}), $\varrho_{\mathrm{gr}}$ is the density of particles, g$_{\mathrm{c}}$ is the gravitational acceleration of the cometary nucleus (g$_{\mathrm{c}}$ = 2.38$\cdot$10$^{-3}$), and P is the period of its rotation (P = 57$\pm$1 h, \citet{Knight2024}). Note that in Eq.(\ref{A2}), to simplify it, the influence from solar radiation pressure has been omitted, which is discussed in detail in \citep{Crifo2005}. The results of the particle radius calculations are shown in Figs.(\ref{A_1}-\ref{A_2}). 
\begin{figure}
\begin{center}
\includegraphics[width=8.5cm]{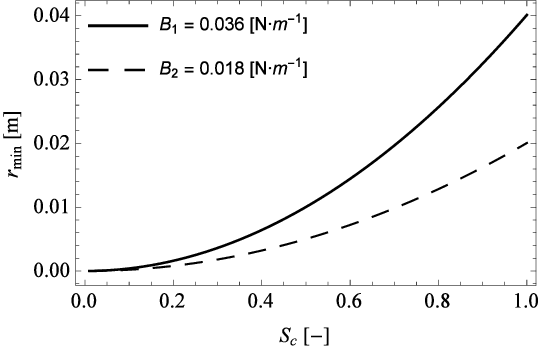}
\end{center}
\caption{The minimum radius of a particle that can be lifted into the coma from the surface of the cometary nucleus as a function of the degree of particle cleanliness (S$_{\mathrm{c}}$). The calculations were performed assuming that comet 12P is at perihelion and the sublimation of water ice controls its activity.}
\label{A_1}
\end{figure}
\begin{figure}
\begin{center}
\includegraphics[width=8.5cm]{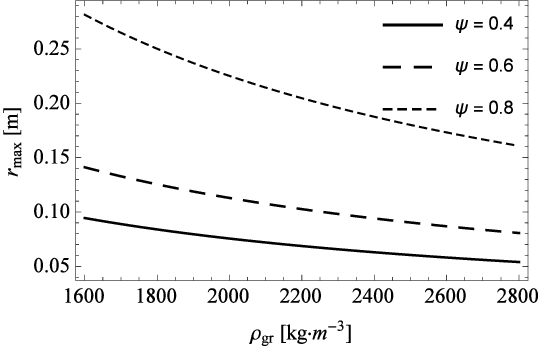}
\end{center}
\caption{The maximum radius of dust agglomerates that can be lifted into the coma from the surface of the cometary nucleus as a function of the particle density ($\varrho_{\mathrm{gr}}$). The calculations are based on the same assumptions as in Fig.(\ref{A_1}).}
\label{A_2}
\end{figure}

\noindent Analyzing the results presented in this paper, it can be noted that cometary outbursts, due to their energetic nature, are responsible for the occurrence of small grains of primary matter in the coma. In contrast, sublimation activity is responsible for the emission of much larger porous dust agglomerates.

\bsp	
\label{lastpage}
\end{document}